\documentclass[doublecol]{epl2}
\usepackage[latin1]{inputenc}
\usepackage{graphicx}
\usepackage{amssymb}
\usepackage{amsmath}
\usepackage{xspace}
\usepackage{dcolumn}
\usepackage{bm}
\usepackage{times}

\title{Metallicity in the half-filled Holstein-Hubbard model}
\shorttitle{Metallicity in the half-filled Holstein-Hubbard model} 
\author{H. Fehske\inst{1}\thanks{E-mail: 
\email{holger.fehske@physik.uni-greifswald.de}} \and G. Hager\inst{2} \and E. Jeckelmann\inst{3}}
\shortauthor{H. Fehske \etal}

\institute{                    
  \inst{1} Institute for Physics, 
Ernst-Moritz-Arndt University Greifswald, D-17487
Greifswald, Germany\\
  \inst{2} Computing Center, Friedrich-Alexander 
University Erlangen-N\"urnberg, D-91058 Erlangen,
Germany\\
\inst{3} Institute for Theoretical Physics, 
Leibniz University Hannover, D-30167 Hannover,
Germany
}

\abstract{
We re-examine the Peierls insulator to Mott insulator transition 
scenario in the one-dimensional Holstein-Hubbard model where, at half-filling, 
electron-phonon and electron-electron interactions compete
for establishing charge- and spin-density-wave states, respectively.
By means of large-scale density-matrix renormalization group calculations
we determine the spin, single-particle and two-particle 
excitation gaps and prove---in the course of a careful finite-size 
scaling analysis---recent claims for an intervening metallic
phase in the weak-coupling regime. We show that for large phonon 
frequencies the metallic region is even more extended than 
previously expected, and subdivided into ordinary Luttinger 
liquid and bipolaronic liquid phases.   
} 

\begin{document}
\maketitle

The challenge of understanding the subtle interplay of 
electron-electron and electron-phonon interaction effects
in low-dimensional condensed matter systems, such as conjugated polymers, 
charge transfer salts, inorganic spin-Peierls compounds,
halogen-bridged transition metal complexes,  
ferroelectric perovskites,  or organic 
superconductors,~\cite{IYS73,TNYS90,BS93,HTU93} has stimulated intense 
work on generic fermion/spin-boson models. 
In this respect the one-dimensional (1D) Holstein\footnote{The Holstein 
model (cf. T. Holstein, Ann. Phys. (N.Y.) {\bf 8}, 325 (1959); {\it ibid.}
{\bf 8}, 343 (1959))  has been studied extensively as a paradigmatic 
model for polaron formation in the low-density limit. For commensurate 
band fillings the coupling to the lattice supports charge ordering.}-Hubbard\footnote{The Hubbard model (cf. J. Hubbard, Proc. Roy. Soc. London, Ser. 
A {\bf 276}, 238 (1963)), originally designed to describe 
ferromagnetism of transition metals, has more recently
been used as the probably most simple model to account for strong 
Coulomb correlation effects in a great variety of materials.}  model (HHM) is particularly rewarding 
to study.~\cite{FWWGBB02,TC03,FWHWB04,CH05,TAA05,NZWL06,TAA07,HC07,MTM06,TTCC07} 
It accounts for a tight-binding electron band, 
an intra-site Coulomb repulsion between electrons of opposite spin,
a local coupling of the charge carriers to optical phonons,
and the energy of the phonon subsystem in harmonic approximation:  
\begin{eqnarray}
 H &=&  - t
      \sum\limits_{\langle i, j\rangle\sigma}\! c_{i\sigma}^\dagger 
      c_{j\sigma}^{} + U
      \sum\limits_i n_{i\uparrow} n_{i\downarrow}\nonumber\\
      &&- \sqrt{\varepsilon_{\rm p}  \omega_0}
      \sum\limits_{i\sigma} (b_i^\dagger + b_i^{}) n_{i\sigma}^{}
      + \omega_0 \sum\limits_i b_i^\dagger b_i^{} \,.
\label{hhm}
    \end{eqnarray} 
In Eq.~(\ref{hhm}), $n_{i\sigma}=c^{\dagger}_{i\sigma}c^{}_{i\sigma}$, where 
$c^{\dagger}_{i\sigma}$ ($c^{}_{i\sigma}$) creates (annihilates)
a spin-$\sigma$ electron at Wannier site $i$ of an 1D lattice 
with $N$ sites, and $b^{\dagger}_{i}$ ($b^{}_{i}$) 
are the corresponding bosonic operators for a dispersionless 
phonon with frequency $\omega_0$.

The physics of the HHM is governed by three 
competing effects: the itinerancy of the electrons ($\propto$\,$t$), 
their on-site Coulomb repulsion ($\propto$\,$ U$), and the 
local electron-phonon (EP) coupling ($\propto$\,$\varepsilon_{\rm p}$). 
Since the EP interaction is retarded, the phonon frequency ($\omega_0$) 
defines a further relevant energy scale. Hence one is advised to introduce 
besides the adiabaticity ratio, 
\begin{equation}
\label{adratio}
\alpha=\omega_0/t\,,
\end{equation}
two dimensionless coupling constants:
\begin{equation}
\label{parratio}
u=U/4t, \quad\mbox{and}\quad \;g^2=\varepsilon_{\rm p}/\omega_0
\quad\mbox{ or} \quad\lambda=\varepsilon_{\rm p}/2t\,.
 \end{equation} 

\begin{figure}[t]
\onefigure[width=0.48\textwidth]{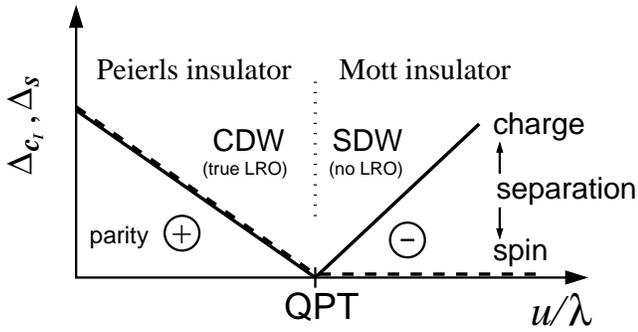}
  \caption{\label{fig_spd}  Schematic representation of the 
Peierls insulator to Mott insulator quantum phase transition 
at fixed $\lambda$ and $\alpha$ in the strong coupling 
small-to-intermediate phonon-frequency regime 
of the 1D half-filled Holstein-Hubbard model. 
The PI exhibits long-range order (LRO) because a discrete
symmetry is broken; the SDW MI has continuous symmetry, 
i.e. there is no LRO in 1D.  $\Delta_s$ (dashed line) 
and $\Delta_{c_1}$ (solid line) 
denote the spin and charge excitation gap, respectively. 
For finite periodic chains with $N=4n$, the transition could be 
identified by a ground-state level crossing associated 
with a change in the parity eigenvalue $P=\pm 1$, where the
site inversion symmetry operator $P$ is defined by 
$Pc_{i\sigma}^\dagger P^\dagger=c_{N-i+1\sigma}^\dagger$ 
(cf. Refs.~\cite{GST00,FWWGBB02}).}
\end{figure}

Both Holstein and Hubbard interactions tend to immobilize the 
charge carriers, and even may drive a metal-insulator transition 
at commensurate band fillings. For the half-filled band case
(one electron per lattice site), most previous analytical and 
numerical studies of the HHM reveal that Peierls insulator (PI) 
or Mott insulator (MI) states are favored over the metallic state
at zero temperature. Whereas the PI is characterized by 
a distortion of the underlying lattice
accompanied by dominant charge-density-wave (CDW) correlations,
the Mott insulator is basically a spin-density-wave (SDW) 
state without any lattice dimerization. 
The physical excitations differ accordingly:  
while ``normal'' electron-hole excitations are expected in the  
PI phase, charge (spin) excitations are known to be 
massive (gapless) in the MI state, at least for the 1D Hubbard-only 
model. Thus, varying the control parameter $u/\lambda$, a cross-over from 
standard quasi-particle behavior to spin-charge separation might 
be observed in the more general HHM (see Fig.~\ref{fig_spd}). 
This scenario has been corroborated by extensive Lanczos diagonalization and 
density-matrix renormalization 
group (DMRG) studies.~\cite{FWWGBB02,FWHWB04,FJ06} 
In particular the existence of a single quantum 
critical point, separating Peierls- and 
Mott-insulating phases at $u/\lambda\simeq 1$, 
has been confirmed by up-to-date stochastic series
expansion (SSE) quantum Monte Carlo (QMC) calculations.~\cite{HC07}
But this only holds in the {\it strong-coupling} and {\it 
adiabatic-to-intermediate phonon-frequency} regime. 

For the pure (spinless and spinful) Holstein model it is well known 
that quantum phonon fluctuations may destroy the Peierls phase,
provided the EP interaction is not too 
strong.~\cite{WHS95,BMH98,JZW99,HWBAF06,CSC05,FWHWBB05}
In this case, a Luttinger liquid phase exists below a critical
coupling $g_c$, where $\lambda_c\to 0$ as $\alpha\to 0$.\footnote{The 
metal-insulator transition at $g_c$ is expected to be 
of Kosterlitz-Thouless type, at least for 
$\omega_0\to \infty$ (cf. Refs.~\cite{HF83}, and~\cite{BMH99}).}  
For this reason, it is natural to examine the stability of 
such a metallic state as the Coulomb interaction is turned on.  
(We call metal any phase with gapless charge excitations.)
An intervening metallic phase would of course prevent a direct 
insulator-to-insulator transition. Recent numerical investigations
of the 1D HHM, based on variable-displacement Lang-Firsov~\cite{TC03},
DMRG~\cite{TAA05,TAA07}, and SSE QMC~\cite{CH05,HC07} approaches, 
give strong evidence that the CDW-SDW transition does indeed 
split into two subsequent CDW-metal and metal-SDW transitions 
in the {\it weak-coupling}  {\it intermediate-to-large 
phonon-frequency} regime.\footnote{A intermediate metallic phase 
has been shown to exist also for the $D=\infty$ HHM 
(see, e.g., Ref.~\cite{WM07} and references therein).} 
To map out the phase diagram is technically
challenging, however, since the energy scales are not well separated
in this region. Furthermore, it is not obvious which physical quantities 
are most suitable to distinguish the various phases. In this respect,
local magnetic moment and effective electronic hopping 
integral~\cite{TC03}, charge-, spin- and (superconducting)
pairing-correlation functions~\cite{TAA05}, or Luttinger 
liquid parameter~\cite{CH05} and charge/spin susceptibilities~\cite{HC07} 
have been proposed.
\begin{figure}[t]
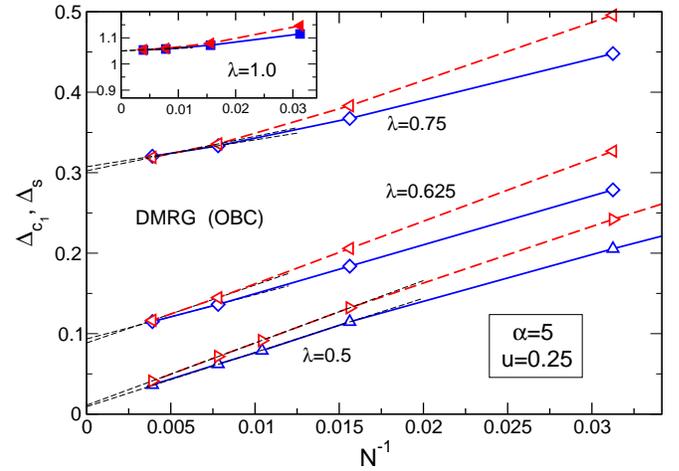

  \onefigure[width=0.48\textwidth]{fig2_cg_sg_aa.eps}
  \caption{\label{fig_cg_sg_aa} (Color online) 
Finite-size scaling of spin (left and right triangles; red dashed lines) 
and charge (diamonds, triangles up, circles; blue solid lines)  excitation 
gaps in the 1D HHM with OBC. Filled (open) symbols belong to 
insulating (metallic) states, see Fig.~\ref{fig_pd_aa}.}  
\end{figure}

In this work our analysis of the weak-to-intermediate
coupling regime of the 1D HHM is based on the behavior of various 
spin and charge excitation gaps. 
These many-body gaps can be calculated by DMRG with high precision 
on a sequence of sufficiently large lattices. 
This allows for a reliable finite-size scaling of the gap data
using their expected asymptotic behavior 
$\Delta(N) = \Delta(\infty) + b/N + c/N^2$ for $N \gg 1$. 
A detailed description of our fermion-boson DMRG pseudo-site method 
(which treats electron and phonon degrees of freedom on an equal footing
in the whole parameter regime) can be found in Ref.~\cite{JF07}.
In the numerical work we use typically 3--5 bosonic pseudosites
and employ open boundary conditions (OBC).   

The charge and spin excitation gaps are determined from 
\begin{equation}
 \label{cg}
 \Delta_{c_1}=E_0^{+}(1/2)+E_0^{-}(-1/2)-2E_0(0)
\end{equation} 
and
\begin{equation}
\label{sg}
 \Delta_{s}=E_0(1)-E_0(0)\,,
 \end{equation}
respectively, where $E_0^{(\pm)}(S^z)$ is the ground-state 
energy at (away from) half-filling with $N_{e}=N$ ($N_{e}=N\pm 1$) 
particles in the sector with total spin-$z$ component $S^z$. 
In the various phases expected to occur in the 1D HHM
$\Delta_{s}$ corresponds to the energy of the lowest spin excitation 
(spin gap). 
The one-particle gap $\Delta_{c_1}$ 
yields the energy of the lowest charge excitation (charge gap) in
the Peierls band insulator, Luttinger liquid, and Mott insulator 
phases and is an upper
limit for this charge gap in any other phase. 
Note that since we compare ground-state energies when calculating the charge 
and spin gaps, lattice relaxation effects arising from different 
particle numbers are included.

Searching for metallicity in the HHM, we first consider 
the {\it anti-adiabatic regime} ($\alpha=5$), because large phonon 
frequencies will clearly act against any static Peierls ordering.
Figure~\ref{fig_cg_sg_aa} shows the scaling of $\Delta_{c_1}$ and 
$\Delta_{s}$ with system size $N$ as the EP coupling $\lambda$ is 
lowered at fixed Coulomb interaction $u=0.25$. 
For large $\lambda$ there is a CDW made up of ``singlet bipolarons'',
i.e. a bipolaronic superlattice.
Since the bipolarons that emerge are rather small objects,
the finite-size dependencies of $\Delta_{c_1}$ and $\Delta_s$ are 
weak. This is just as in the strong-coupling 
non-too-anti-adiabatic regime~\cite{Ha05} 
(cf. also Fig.~11 in Ref.~\cite{FJ06}). With decreasing EP coupling,
i.e. increasing ratio $u/\lambda$, both $\Delta_{c_1}$ and $\Delta_s$
become smaller, acquire a notable finite-size dependency,
and finally both scale to zero, indicating Luttinger-liquid 
metallic behavior. Since $\Delta_{c_1}$ and $\Delta_s$ converge to the same value 
as $N\to\infty$, spin and charge degrees of freedom ``stick together''.
Above a critical ratio $u/\lambda$, however, a SDW (Mott insulating) phase 
is realized. In the MI we found a finite charge excitation gap, which
in the limit $\lambda/u\ll 1$ scales to the charge gap 
of the Hubbard model, whereas the extrapolated spin gap 
remains zero (cf. the data for $\lambda=0.05$). 
This is in agreement with spin-charge separation.

\begin{figure}[t]
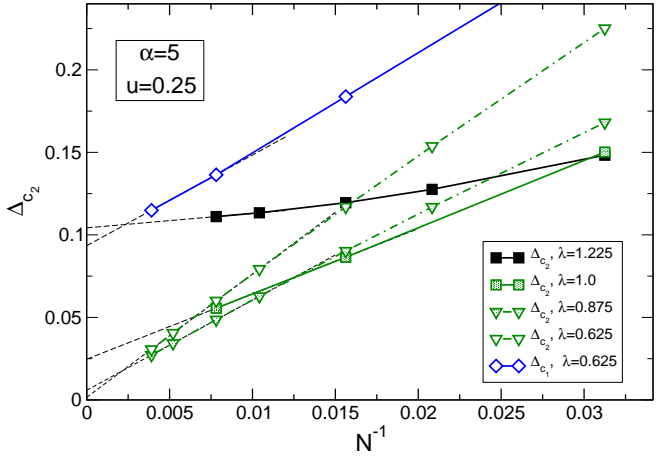

  \onefigure[width=0.48\textwidth]{fig3_c2g_aa.eps}
  \caption{\label{fig_c2g_aa} (Color online) Finite-size scaling of 
the two-particle excitation gap $\Delta_{c_2}$ in the HHM. 
For comparison we include $\Delta_{c_1}$ in the bipolaron liquid  
phase. For further explanation see text.}
\end{figure}
In order to gain deeper insight into the nature of the intermediate phase
between PI and MI we determine, besides $\Delta_{c_1}$, the two-particle
excitation gap  
\begin{equation}
 \label{tpg}
 \Delta_{c_2}=E_0^{2+}(0)+E_0^{2-}(0)-2E_0(0)\,.
 \end{equation}
This gap corresponds to the charge gap in a bipolaronic insulator and
is an upper limit for it in any other phase. 
Of course, one- and two-particle excitation gaps should simultaneously 
open if we enter the PI and MI phases. If the PI phase is
a bipolaronic insulator (superlattice) rather than a traditional
Peierls band insulator, mobile bipolarons may occur first in the 
dissolving process of the PI, as the $\lambda/u$ ratio is lowered.
Such a bipolaronic metal/liquid phase will then be characterized 
by $\Delta_{c_2}=0$ but finite $\Delta_{c_1}$ (and $\Delta_s$). 
Adding/removing a single particle from the metallic bipolaron phase   
is energetically costly because the bipolarons are (tightly) bound.
A bipolaron as a whole, however, can be added or removed without effort.     

Figure~\ref{fig_c2g_aa} illustrates that this scenario holds in 
the anti-adiabatic weak-coupling regime. As the EP coupling 
gets weaker, we enter a region where $\Delta_{c_1}>0$ but $\Delta_{c_2}\to 0$ 
(see, e.g., the data for $\lambda=0.625$). $\Delta_{c_2}$ stays zero 
as $\Delta_{c_1}$ vanishes at still smaller $\lambda$ ($u$ fixed),
until we enter the MI state. 

\begin{figure}[t]
  \includegraphics[width=0.48\textwidth]{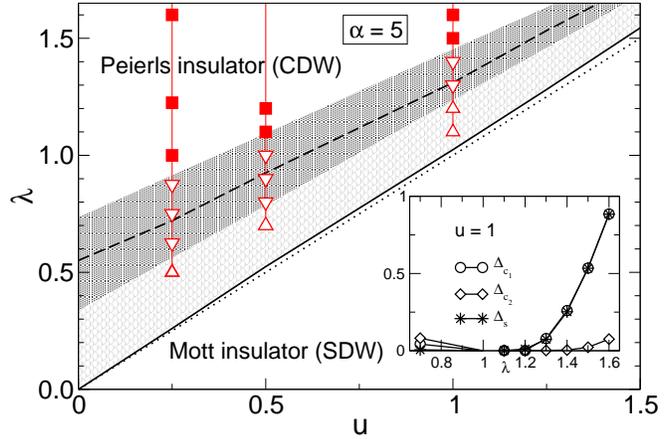}
  \caption{\label{fig_pd_aa}  (Color online) Phase diagram of the 
half-filled Holstein-Hubbard model in the weak-coupling anti-adiabatic 
($\alpha=5$) regime.
Here filled squares, open triangles down, and open triangles up,
denote the PI, bipolaronic metal, and Luttinger-liquid metal phases, 
respectively.
The extension of the two latter phases is marked by the intensely
(weakly) shaded regions. Dashed (solid) lines designate the 
Peierls--intermediate 
state (intermediate phase--Mott) phase boundaries obtained by 
Ref.~\cite{CH05}. The dotted line indicates $u=\lambda$.
The inset gives the  ($N\to \infty$) extrapolated values of 
the one-particle, two particle, and spin excitation gaps at $u=1$. Note that
$\Delta_{c_2}$ is twice as large as $\Delta_{c_1}$ in the MI phase.}
\end{figure}

Figure~\ref{fig_pd_aa} combines the findings of our scaling
analysis for $\Delta_s$, $\Delta_{c_1}$ and $\Delta_{c_2}$
with the results of previous studies of the 
half-filled HHM based on susceptibility and Luttinger 
liquid parameter data~\cite{CH05,HC07}. 
First, we note that the PI-metal phase boundary is shifted
compared to the results of Ref.~\cite{CH05}, 
while the metal-MI transition line is the very same. 
That is, taking $\Delta_{c_2}=0$ as a criterion 
for the instability of the PI phase, we find an even larger 
region for the PI-MI intervening state. Second,  within the metallic state,
our data suggests a cross-over between a bipolaronic liquid ($\Delta_{c_2}=0$; 
$\Delta_{c_1}$, $\Delta_{s}$ $>0$) and a Luttinger liquid
($\Delta_{c_2}=\Delta_{c_1}=\Delta_{s}=0$).\footnote{There is an ongoing 
discussion, whether superconducting pairing correlations may become
dominant in the HHM between the CDW and SDW phases (cf., e.g. 
Ref.~\cite{TAA05}). Here, we will not address the highly controversial 
issue of existence of superconducting ground states 
(off-diagonal long-range order) in a stricly 1D electron-phonon model.}  

\begin{figure}[t]
  \includegraphics[width=0.4\textwidth]{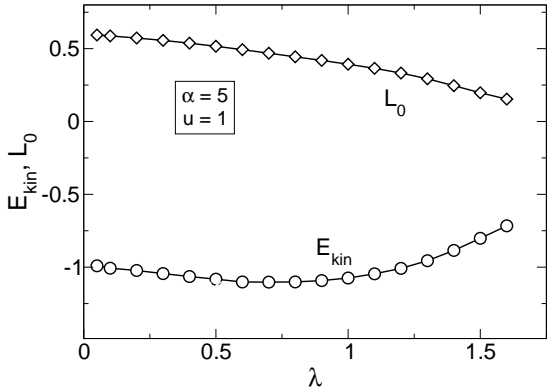}
  \caption{\label{fig_l0ekin_aa}  Kinetic energy $E_{kin}$ 
and local magnetic moment $L_0$ at $u=1$, $\alpha=5$.}
\end{figure}

The mean kinetic energy,  
\begin{equation}
E_{kin}=-t \sum_{\langle i, j\rangle\sigma}\langle c_{i\sigma}^\dagger 
      c_{j\sigma}^{}\rangle\,,
\end{equation}
shown in Fig.~\ref{fig_l0ekin_aa} for $u=1$,   
exhibits a pronounced minimum in the intermediate metallic 
phase~\cite{TC03}. The local magnetic moment, 
\begin{equation}
L_0=\frac{3}{4N}\sum_i
\langle (n_{i,\uparrow}-n_{i,\downarrow})^2\rangle\,,
\end{equation}
can be taken as as a measure for the ``localization'' of the electron spin.
For the pure half-filled Hubbard model it varies between 3/8 (band limit)
and 3/4 (atomic limit). As can be seen from Fig.~\ref{fig_l0ekin_aa},
$L_0$ is suppressed in the PI CDW phase because of the alternating 
arrangement of (almost) empty and double occupied sites. Entering 
the metallic phase $L_0$ is enhanced and reaches its maximum in the 
MI phase. Both $E_{kin}$ and $L_0$ are however not very suitable
for fixing the phase boundaries.  

\begin{figure}[t]
  \includegraphics[width=0.46\textwidth]{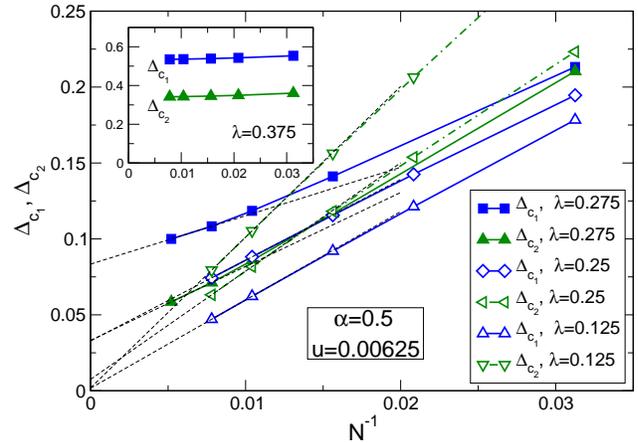}
  \caption{\label{fig_c2g_cg_a}  (Color online) Finite-size scaling
of the one- and two-particle excitation gaps in the adiabatic ($\alpha=0.5$)
regime of the HHM. The inset gives the results in the PI phase 
$\lambda=0.375$.}
\end{figure}

\begin{figure}[h!]
  \includegraphics[width=0.48\textwidth]{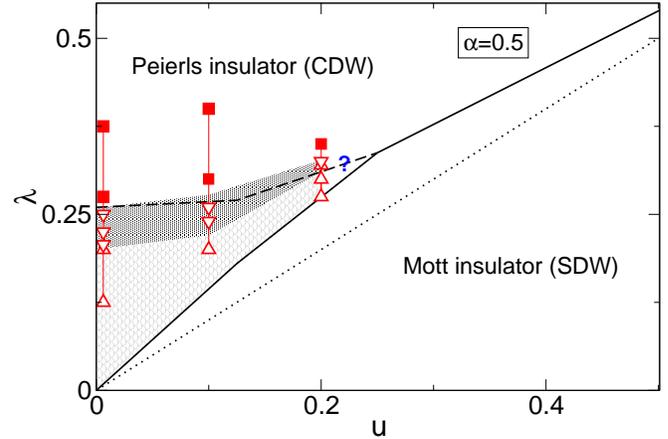}
  \caption{\label{fig_pd_a} (Color online)  Phase diagram of the HHM 
in the adiabatic weak-coupling  regime. Again filled squares, 
open triangles down, and open triangles up, denote the PI, 
bipolaronic metal, and Luttinger-liquid metal phases, 
respectively. Dashed (solid) lines are taken 
from Ref.~\cite{CH05}; the dotted line gives $u=\lambda$.
Since the finite-size scaling becomes exceedingly costly
in the vicinity of the tricritical point $u\simeq 0.25, \lambda=0.3375$
given by Clay et al.~\cite{CH05}, we cannot reliably resolve the 
phase structure there.}
\end{figure}

The corresponding results for the {\it adiabatic regime} ($\alpha=0.5$) 
are given in Figs.~\ref{fig_c2g_cg_a} and~\ref{fig_pd_a}.
Again we have strong evidence for an intermediate metallic state. 
The region where bound mobile charge carriers (bipolarons) exist,
however, now is a small strip between the PI and metal phases only, 
and expected to vanish if the adiabaticity 
ratio $\alpha$ goes to zero.\footnote{For the so-called adiabatic
HHM with frozen phonons an additional bond-order-wave might occur 
within the intermediate phase.~\cite{FKSW03}}   
If all energy scales---set by $\lambda$, $u$, $\alpha$---become small, 
the finite-size scaling is extremely delicate and it is difficult
to resolve the different metallic phases. The main differences   
in comparison to the anti-adiabatic regime pertain to (i) the nature 
of the Peierls phase, which is a traditional band insulator 
in the adiabatic case, and (ii) the existence of a tri-critical 
point at moderate coupling strengths.

To summarize, we complemented previous numerical 
investigations of the half-filled Holstein-Hubbard 
model~\cite{TC03,FWHWB04,TAA05,TAA07,CH05,HC07} by 
a large-scale DMRG analysis of the weak electron-phonon 
and electron-electron interaction regime.   
By calculating spin and charge excitation gaps we confirmed 
the existence of an intermediate metallic phase in the 
cross-over region of the Peierls-insulator Mott-insulator 
transition. This phase is shown to be most extended in the 
anti-adiabatic limit of large phonon frequencies.
If the PI typifies a bipolaronic superlattice,
it gives way to a bipolaronic liquid composed of bound 
mobile (singlet) charge carriers as the EP coupling weakens.  
Reducing the EP to Coulomb interaction ratio further, a
cross-over to a metallic state with unbound (weakly 
phonon-dressed) electrons (Luttinger liquid) takes place, until the MI state
is finally reached at $\lambda \lesssim u$. If the PI typifies a 
rather standard band insulator in the adiabatic, small 
phonon-frequency limit, the bipolaron liquid phase narrows 
substantially, but still a metallic phase exists in between 
CDW and SDW states. Putting these findings together with
the results previously obtained for the strong-coupling case, 
where a direct first-order PI-MI transition takes place,
a rather complete picture of the physical properties of the 
1D HHM emerges. 

{\it Acknowledgments.}
We would like to thank A. Alvermann, F. F. Assaad, K. W. Becker,  
S. Ejima, M. Hohenadler, and G. Wellein for valuable discussions. 
This work was supported by DFG through SFB 652 and KONWIHR Bavaria 
project HQS@HPC. Furthermore, we acknowledge generous computer 
time grants by LRZ Munich and HLRN Berlin. 

\end{document}